\documentclass[
    ,final            
  ,numberedheadings 
  ]
  {aipproc}

\usepackage[mathscr]{eucal}
 \usepackage{amsmath, amssymb}

  \def\sun{\hbox{$\odot$}}
   \def\degr{\hbox{$^\circ$}}
\newcommand{\be}{\begin{equation}}
\newcommand{\ee}{\end{equation}}
\newcommand{\bdm}{\begin{displaymath}}
\newcommand{\edm}{\end{displaymath}}
\def\la{\mathrel{\mathchoice {\vcenter{\offinterlineskip\halign{\hfil
$\displaystyle##$\hfil\cr<\cr\sim\cr}}}
{\vcenter{\offinterlineskip\halign{\hfil$\textstyle##$\hfil\cr
<\cr\sim\cr}}}
{\vcenter{\offinterlineskip\halign{\hfil$\scriptstyle##$\hfil\cr
<\cr\sim\cr}}}
{\vcenter{\offinterlineskip\halign{\hfil$\scriptscriptstyle##$\hfil\cr
<\cr\sim\cr}}}}}

\def\ga{\mathrel{\mathchoice {\vcenter{\offinterlineskip\halign{\hfil
$\displaystyle##$\hfil\cr>\cr\sim\cr}}}
{\vcenter{\offinterlineskip\halign{\hfil$\textstyle##$\hfil\cr
>\cr\sim\cr}}}
{\vcenter{\offinterlineskip\halign{\hfil$\scriptstyle##$\hfil\cr
>\cr\sim\cr}}}
{\vcenter{\offinterlineskip\halign{\hfil$\scriptscriptstyle##$\hfil\cr
>\cr\sim\cr}}}}}
\layoutstyle{6x9}

\begin{document}

\noindent {\it to appear in Astronomy Reports, Vol.~57, March 2013}

\vspace{1cm}

\title{On the spin-down mechanism of the X-ray pulsar 4U~2206+54}

\classification{97.10.Gz, 97.80.Jp, 95.30.Qd}
\keywords{Accretion, X-ray binaries, neutron star, pulsars, magnetic field}

\author{N.R.\,Ikhsanov}{
  address={Pulkovo Observatory, Pulkovskoe Shosse 65, Saint-Petersburg 196140, Russia}
}

\author{N.G.\,Beskrovnaya}{
  address={Pulkovo Observatory, Pulkovskoe Shosse 65, Saint-Petersburg 196140, Russia}
}

\begin{abstract}
Observations of the X-ray pulsar 4U~2206+54, performed over a 15-year span, show its period, which is now $5555\pm9$\,s, to
increase drastically. Such behavior of the pulsar can be hardly explained in the frame of traditional scenarios of spin evolution of
compact stars. The observed spin-down rate of the  neutron star in 4U~2206+54 proves to be in good agreement with the value expected
within magnetic accretion scenario which takes into account that magnetic field of the accretion stream can under certain conditions
affect its geometry and type of flow.  The neutron star in this case accretes material from a dense gaseous slab with small angular
momentum which is kept in equilibrium by the magnetic field of the flow itself. The magnetic accretion scenario can be realized in
4U~2206+54 provided the magnetic field strength on the surface of the optical counterpart to the neutron star is in excess of 70\,G.
The magnetic field strength on the surface of the neutron star in the frame of this scenario turns out to be $4 \times 10^{12}$\,G
which is in accordance with estimates of this parameter from the analysis of X-ray spectra of the pulsar.
\end{abstract}

\maketitle


  \section{Introduction}

4U~2206+54 is a moderate-luminosity X-ray source discovered at the beginning of X-ray astronomy \cite{Giacconi-etal-1972}. It has been identified with a neutron star in a high-mass binary with the orbital period $P_{\rm orb} \simeq 19.25$\,d \cite{Corbet-etal-2007} situated at the distance of about  2.6\,kpc \cite{Blay-etal-2006}. Its optical counterpart is the O9.5V star  BD~+53$^{\degr}$\,2790 underfilling its Roche lobe and losing mass via relatively slow, $v_{\rm w} \sim 300-500\,{\rm km\,s^{-1}}$, dense stellar wind \cite{Ribo-etal-2006, Blay-Reglero-2011}. The neutron star rotates with the period $P_{\rm s} = 5555\pm9$\,s
\cite{Rieg-etal-2009, Finger-etal-2010}. The X-ray spectrum of the system is typical for a neutron star accreting material onto its
magnetic poles. The average X-ray luminosity throughout the entire duration of observations ranges within  $L_{\rm x} \sim (2-4) \times 10^{35}\,{\rm erg\,s^{-1}}$. The individual X-ray light curves exhibit chaotic variations of relatively low amplitude on a
timescale of 30 minutes. Higher-amplitude fluctuations of 15 to 45 hours duration and a peak luminosity of $\sim 4 \times 10^{36}\,{\rm erg\,s^{-1}}$ were detected in hard X-rays with INTEGRAL mission \cite{Wang-2010}.

Analyzing X-ray light-curves of  4U~2206+54 obtained in  1998--2007 Finger et al.  \cite{Finger-etal-2010} reported the spin frequency of the neutron star  ($\nu = 1/P_{\rm s}$) to decrease at the rate $\dot{\nu}_{\rm sd} \simeq (-1.7 \pm 0.7) \times 10^{-14}\,{\rm Hz\,s^{-1}}$. The characteristic spin-down time-scale of the neutron star, $\tau_{\rm sd} = P_{\rm s}/2\,\dot{P} \sim 180$\,yr, in this case is very short compared to a typical age of neutron stars in the high-mass X-ray binaries ($\ga 10^6-10^7$\,yr, see, e.g. \cite{Masevich-Tutukov-1988}). This testifies the episodic character of observed variations of the spin period and allows to assume that the star rotates near its equilibrium period. It is also worth mentioning that the X-ray luminosity of the system during this time remained close to its average value with except for hard X-ray flares detected at the end of 2005 \cite{Wang-2010}. This indicates an absence of direct correlation between variations of the star's spin period and mass accretion rate onto its surface and marks it possible to discard a possibility to describe the observed spin evolution exclusively in terms of variations of the spin-up torque applied to the star from the accretion flow (for discussion see \cite{Finger-etal-2010}).

Interpretation of the observed spin-down  rate of the neutron star in the frame of popular accretion scenarios was discussed in
\cite{Finger-etal-2010, Ikhsanov-Beskrovnaya-2010}. In particular, the authors argued that a presence of the Keplerian accretion disk is the system is hardly probable. The spin-down of the star in this case could occur only under condition that the magnetic field
strength on its surface exceeds  $10^{15}$\,G. In order to explain the observed spin-down rate of the star within the picture  of
free-falling quasi-spherical accretion flow it is necessary to assume that the value of the surface magnetic field of the neutron
star is in excess of  $10^{14}$\,G. Finally, it was shown that even within the model of accretion from a turbulent spherical envelope the observed behavior of the pulsar can be described only provided the magnetic field strength on the surface of the compact star is at least  $6 \times 10^{13}$\,G and, thus, surpasses the  quantum critical limit to the magnetic field strength  $B_{\rm Q} = m_{\rm e}^2c^3/e \hbar \simeq 4.4 \times 10^{13}$\,G. The neutron star in the frame of this scenario turns out to be an
accreting magnetar of the age $\la 30\,000$\,yr estimated from the characteristic decay time of supercritical magnetic field \cite{Ikhsanov-Beskrovnaya-2010}.

An assumption about supercritical value of magnetic field in neutron stars seems quite plausible in the light of investigation of
anomalous X-ray pulsars and soft gamma-ray repeaters. However, it is not obvious that this assumption is acceptable in the case of
long-period X-ray pulsars.  In particular, analysis of the X-ray spectrum of  4U~2206+54, presented in  \cite{Torrejon-etal-2004},
has shown that surface magnetic field of the neutron star in this system is $B_{\rm CRSF} \simeq 4 \times 10^{12}$\,G. This is
testified by the observed energy distribution in the spectrum of this source and a presence of absorbtion feature in the region of
30\,keV identified by the authors with a cyclotron line. It should be noted that in case of relatively slow stellar wind,  $v_{\rm w} \sim 300-500\,{\rm km\,s^{-1}}$, and low rate of mass exchange between the system components,
 \be
\dot{M}_{\rm a} \sim 2 \times 10^{15}\ m\ R_6 \left(\frac{L_{\rm X}}{4 \times 10^{35}\,{\rm erg\,s^{-1}}}\right)\ {\rm g\,s^{-1}},
 \ee
the spin-down time-scale of the star in the propeller state (see equation~(21) in  \cite{Ikhsanov-2007}),
  \be\label{tauc}
\tau_{\rm c} \ga 10^5\ I_{45}\ \mu_{32}^{-1}\ \dot{M}_{15.6}^{-1/2}\ \left(\frac{v_{\rm w}}{400\,{\rm km\,s^{-1}}}\right)^{-3/2}\
{\rm yr},
 \ee
exceeds the decay timescale of supercritical magnetic field  (see \cite{Colpi-etal-2000, Bonanno-etal-2006}). Here $m$, $R_6$ and
$I_{45}$ are the mass, radius and the moment of inertia of the neutron star expressed in units of  $1.4\,M_{\sun}$, $10^6$\,cm and
$10^{45}\,{\rm g\,cm^2}$, respectively, while $\mu_{32}$ is the dipole magnetic moment of the star in units of $10^{32}\,{\rm  G\,cm^3}$. $\dot{M}_{15}= 10^{15}\,{\rm g\,s^{-1}}$ is the rate of mass exchange between the system components which in the case of stationary accretion equals the mass accretion rate onto the neutron star surface. This result raises doubts concerning the validity of the assumption about extremely high magnetic field of this object.

A situation in which the magnetic field strength obtained in the spin evolution models significantly exceeds the value estimated from the cyclotron line measurements is not unique. For instance, such discrepancy was revealed for a long-period X-ray pulsar  GX\,301--2 \cite{Lipunov-1982, Doroshenko-etal-2010}. Our studies \cite{Ikhsanov-etal-2012, Ikhsanov-Beskrovnaya-2012} have shown that this inconsistency can be caused by oversimplification of the accretion picture used in the modeling of spin evolution of X-ray pulsars in massive binary systems. In particular, this flaw can be avoided in the frame of magnetic accretion scenario \cite{Shvartsman-1971, Bisnovatyi-Kogan-Ruzmaikin-1974, Bisnovatyi-Kogan-Ruzmaikin-1976} accounting for the influence of the intrinsic magnetic field of the accretion flow on its geometry and parameters. In this paper we show that conditions  necessary for the magnetic accretion to be realized in 4U~2206+54 can be satisfied if the magnetic field strength on the surface of the massive component exceeds 70\,G. The spin-down rate of the neutron star evaluated within this scenario corresponds to its observed value in case the magnetic field strength of the neutron star is of the order of  $\sim B_{\rm CRSF}$.

 \section{Magnetic accretion in 4U~2206+54}

4U~2206+54 is a close binary system with wind-fed accretion. The neutron star which moves through the stellar wind of its massive
companion with the relative velocity $v_{\rm rel}$, captures matter at the rate $\dot{M} = \pi R_{\rm G}^2 \rho_{\infty} v_{\rm rel}$. Here $R_{\rm G} = 2 GM_{\rm ns}/v_{\rm rel}^2$ is the gravitational (Bondi) radius of the neutron star, $M_{\rm ns}$ is its mass and  $\rho_{\infty}$  is the wind density at the Bondi radius. As the captured material is approaching the neutron star under the force of its gravitational attraction it interacts with the stellar magnetic field. This leads to the formation of magnetosphere
of radius $r_{\rm m}$  which is defined by equating the ram pressure of the accretion flow with the magnetic pressure in the neutron star's dipole field. The magnetosphere in the first approximation prevents the accretion flow from reaching the stellar surface.
The flow is decelerating at the magnetospheric boundary and after penetrating the magnetospheric field and moving along the field lines reaches the neutron star surface in the region of magnetic poles.

The radius of magnetosphere depends on the structure of accretion flow beyond its boundary which is determined by the following key parameters: the sound speed,  $c_{\rm s0}$, and the magnetic field strength $B_{\rm f0}$, in the matter captured by the neutron star at the Bondi radius, as well as the relative velocity of the neutron star with respect to the stellar wind of its massive counterpart, $v_{\rm rel}$, and the orbital angular velocity of the binary system  $\Omega_{\rm orb} = 2 \pi/P_{\rm orb}$. Depending on relationship between these parameters the process of accretion proceeds either via a Keplerian disk formation or in the  quasi-spherical or magnetic regime.

  \subsection{Non-magnetic accretion}

In the conventional (non-magnetic) accretion scenarios the magnetic field in the matter captured by the neutron star at the Bondi radius is assumed to be negligible and not to influence the structure of accretion flow. The model of spherical accretion within this approach (a so called Bondi accretion) was first introduced in \cite{Bondi-Hoyle-1944, Bondi-1952} and was further used in the modeling of accretion in the X-ray pulsars \cite{Arons-Lea-1976, Elsner-Lamb-1977}. In the frame of this model the accreting
material is approaching a neutron star in a state of free-fall. Moving at the free-fall velocity $v_{\rm ff} = \left(2GM_{\rm ns}/r\right)^{1/2}$ it reaches the surface of the neutron star at dynamical (free-fall) time-scale $t_{\rm ff} = \left(r^3/2 M_{\rm
ns}\right)^{1/2}$. The ram pressure in the accreting flow is increasing as it approaches the neutron star as \cite{Bondi-1952}
   \be\label{eramr}
 E_{\rm ram}(r) = E_{\rm ram}(R_{\rm G}) \left(\frac{R_{\rm G}}{r} \right)^{5/2},
 \ee
where $E_{\rm ram}(R_{\rm G}) = \rho_{\infty} v_{\rm rel}^2$ is the ram pressure at the Bondi radius. The balance between the ram pressure of the spherical flow and the magnetic pressure of the neutron star dipole field $P_{\rm m} = \mu^2/2 \pi r^6$, is reached at the radius \cite{Arons-Lea-1976}
 \be\label{ra}
 r_{\rm a} = \left(\frac{\mu^2}{\dot{M} (2 GM_{\rm ns})^{1/2}}\right)^{2/7},
 \ee
where $\mu$ is the dipole magnetic moment of the neutron star. This is a magnetospheric radius of the neutron star in the non-magnetic accretion approximation.

Quasi-spherical accretion is a modification of Bondi accretion accounting for a non-zero angular momentum of matter captured by a neutron star in a binary system. The process of mass accretion in this case is accompanied by the angular momentum accretion at the
rate $\dot{J} = \xi\,\Omega_{\rm orb}\,R_{\rm G}^2\, \dot{M}$ (see \cite{Illarionov-Sunyaev-1975} and references therein), where $\xi$ is a parameter accounting for angular momentum dissipation in the accretion flow. The average value of this parameter calculated in the numerical modeling of the wind-fed accretion under assumption about a zero magnetic field of the flow is $<\xi> = 0.2$ (see \cite{Ruffert-1999} and references therein).

Deviation of the quasi-spherical flow geometry from spherical symmetry remains insignificant under condition  $r_{\rm circ} < r_{\rm a}$, where $r_{\rm circ} = \dot{J}^2/GM_{\rm ns} \dot{M}^2$  is a circularization radius, at which the angular velocity of the accreting material,  $\omega_{\rm en} = \xi\,\Omega_{\rm orb}\,\left(R_{\rm G}/r \right)^2$, equals the Keplerian angular velocity, $\omega_{\rm k} = \left(r^3/2GM_{\rm ns}\right)^{1/2}$. Otherwise, a Keplerian accretion disk forms in the system. The condition of disk formation  ($r_{\rm circ} > r_{\rm a}$) can be written as $v_{\rm rel} < v_{\rm cr}$, where \cite{Ikhsanov-2007}
  \be\label{vcr}
v_{\rm cr} \simeq 210\ \xi_{0.2}^{1/4}\ \mu_{30}^{-1/14}\ m^{11/28}\ \dot{M}_{15}^{1/28}\ \left(\frac{P_{\rm orb}}{19\,{\rm d}}\right)^{-1/4}\ {\rm km\,s^{-1}}.
 \ee
Here $\xi_{0.2} = \xi/0.2$. For the parameters of 4U~2206+54 we get the value of $v_{\rm cr}$ significantly less than the stellar wind velocity derived from observations. This means that formation of the Keplerian accretion disk is rather unlikely from the
theoretical point of view. In this light an absence of observational manifestations of the Keplerian disk in the system  seems rather natural  \cite{Torrejon-etal-2004}.

  \subsection{Magnetic accretion}

As first shown by Shvartsman  \cite{Shvartsman-1971}, the picture of wind-fed accretion in the high-mass X-ray binaries can essentially differ from expected within Bondi scenario if the accreting material is magnetized. In the free-falling accretion flow
magnetic field is predominantly radial. This is caused by the fact that in the process of spherical accretion all transverse scales contract as $\sim r^{-2}$ while the longitudinal scales expand as$\sim r^{1/2}$ \cite{Zeldovich-Shakura-1969}. Then under the condition of magnetic flux conservation the field strength in the accreting matter rises as $B_{\rm r} \propto r^{-2}$ \cite{Bisnovatyi-Kogan-Fridman-1970}. This means that the magnetic energy density $E_{\rm m} = B_{\rm r}^2/8 \pi$, in the free-falling plasma,
 \be
 E_{\rm m}(r) = E_{\rm m}(R_{\rm G}) \left(\frac{R_{\rm G}}{r}\right)^4,
 \ee
increases more rapidly than the kinetic energy of its radial motion $E_{\rm ram}(r) \propto \left(R_{\rm G}/r \right)^{5/2}$ (see expression~\ref{eramr}). Here $E_{\rm m}(R_{\rm G}) = \beta^{-1} E_{\rm th}(R_{\rm G})$ is the magnetic energy density in the accretion flow at the Bondi radius normalized using the parameter $\beta$ to the thermal energy of the flow, $E_{\rm th}(R_{\rm G}) = \rho_{\infty} c_{\rm s0}^2$. The distance at which the magnetic energy density in the accretion flow balances its kinetic energy density (the  Shvartsman radius) can be obtained solving the equation  $E_{\rm m}(R_{\rm sh}) = E_{\rm ram}(R_{\rm sh})$, in the form
  \be\label{rsh}
 R_{\rm sh} = \beta^{-2/3}\ \frac{2 GM_{\rm ns} c_{\rm s0}^{4/3}}{v_{\rm rel}^{10/3}}.
 \ee

Shvartsman \cite{Shvartsman-1971} has noted that mass accretion in the spatial region $r < R_{\rm sh}$ can take place as long as the magnetic field in the flow dissipates. Otherwise the magnetic energy in the accretion flow would exceed its total energy (equal to gravitational energy of the star at a given radius) which is impossible under the energy conservation law. Thus, the velocity of radial motion in this region cannot exceed $v_{\rm r} \sim r/t_{\rm rec}$, where
 \be\label{trec}
 t_{\rm rec} = \frac{r}{\eta_{\rm m} v_{\rm A}} = \eta_{\rm m}^{-1}\ t_{\rm ff}\ \left(\frac{v_{\rm ff}}{v_{\rm A}}\right)
 \ee
is the time of field dissipation due to magnetic reconnection, and $v_{\rm A} = B_{\rm r}/(4 \pi \rho)^{1/2}$  is the Alfv\'en velocity in the accretion flow. $\eta_{\rm m}$ is a parameter allowing  for the efficiency of reconnection. Its value depends on both the physical parameters in plasma and configuration of the magnetic field and in general case ranges as  $0 < \eta_{\rm m} <0.1$ \cite{Kadomtsev-1987}. Since  $v_{\rm A} \leq v_{\rm ff}$ (the equality is met at the Shvartsman radius) the time of field dissipation remains significantly less than dynamical timescale, $t_{\rm ff}$, throughout the accretion process. On one hand this justifies an assumption about the magnetic flux conservation in the free-falling plasma, but on the other hand this means that accretion flow is decelerated  by its own magnetic field at the Shvartsman radius and further matter inflow proceeds in the diffusion regime.

Basic conclusions of the scenario proposed by Shvartsman were later confirmed in quantitative modeling by Bisnovatyi-Kogan and Ruzmaikin \cite{Bisnovatyi-Kogan-Ruzmaikin-1974, Bisnovatyi-Kogan-Ruzmaikin-1976}, and by the results of numerical simulations of the spherical magnetic accretion presented in \cite{Igumenshchev-etal-2003, Igumenshchev-2006}. These authors have shown that magnetic field amplification in the free-falling material leads to deceleration of the accretion flow  and its shock-heating up to adiabatic temperature. Accretion inside the Shvartsman radius occurs on the timescale of magnetic field dissipation. The accretion picture in this case depends on the efficiency of cooling processes in the accreting material. If the cooling time at the Shvartsman radius, $t_{\rm cool}(R_{\rm sh})$, exceeds the heating time due to magnetic energy dissipation, $t_{\rm rec}(R_{\rm sh})$, the accretion flow transforms into a hot turbulent envelope with some portion of matter leaving the system in a form of jets \cite{Igumenshchev-etal-2003, Igumenshchev-2006}. Otherwise, the flow forms the magnetic slab whose structure is determined by geometry of the large-scale magnetic field in the accreting matter \cite{Bisnovatyi-Kogan-Ruzmaikin-1974, Bisnovatyi-Kogan-Ruzmaikin-1976}.

In the X-ray pulsars the magnetic field of the accretion flow influences its structure in case  $R_{\rm sh} > r_{\rm a}$. This condition is met if the relative velocity of the neutron star with respect to the wind of its massive companion satisfies inequality
$v_{\rm rel} < v_{\rm mca}$, where \cite{Ikhsanov-Beskrovnaya-2012}
 \be
 v_{\rm mca} = \beta^{-1/5}\,(2GM_{\rm ns})^{12/35}\,\mu^{-6/35}\,\dot{M}^{3/35}\ c_{\rm s}^{2/5}.
 \ee
Substituting parameters of 4U~2206+54, we find
  \be
v_{\rm mca} \simeq 460\ \beta^{-1/5}\ m^{12/35} \mu_{30}^{-6/35}\ \dot{M}_{15}^{3/35}\ \left(\frac{c_{\rm s}}{10\,{\rm km\,s^{-1}}}\right)^{2/5}\ {\rm km\,s^{-1}}.
 \ee
Thus, the accretion process in this system would occur according to magnetic accretion scenario if  $\beta \sim 1$. Estimating the thermal energy density of the matter captured by the neutron star at the Bondi radius,
 \be
 E_{\rm th} \simeq 10^{-4}\,{\rm erg\,cm^{-3}}\ m^{-2}\ \dot{M}_{15}\ \left(\frac{v_{\rm rel}}{400\,{\rm km\,s^{-1}}}\right)^3 \left(\frac{c_{\rm s0}}{10\,{\rm km\,s^{-1}}}\right)^2,
 \ee
and its magnetic energy density,
  \be\label{bwa}
E_{\rm m}(a) \simeq 10^{-3}\ {\rm erg\,cm^{-3}}\ a_{13}^{-4} \left(\frac{\mu_{\rm ms}}{10^{38}\,{\rm G\,cm^3}}\right)^2 \left(\frac{a_{\rm k}}{100\,R_{\sun}}\right)^{-2},
 \ee
we can conclude that the accretion process in 4U~2206+54 should be described in terms of magnetic accretion if the strength of the dipole magnetic field on the surface of the massive (optical) component of this system exceeds  $70$\,G. Here $a_{13} = a/10^{13}$\,cm is the orbital separation, $\mu_{\rm ms}$ is the dipole magnetic moment of the massive star and $a_{\rm k}$ is the distance (measured from the massive component) at which the kinetic energy density of the stellar wind reaches the magnetic energy density of its dipole field (for discussion see \cite{Ikhsanov-Beskrovnaya-2012}). The free-fall of matter captured by the neutron star in this case will be terminated by the intrinsic magnetic field of the accretion flow at the Shvartsman radius exceeding the
magnetospheric radius of the star. The accretion process inside the Shvartsman radius will take place on the timescale of magnetic field dissipation $\sim t_{\rm rec}$.

In order to understand the structure of accretion flow inside the Shvartsman radius it is necessary to take into consideration that the time of its cooling due to inverse Compton scattering of the X-ray photons emitted from the neutron star surface on hot electrons of plasma surrounding its magnetosphere \cite{Elsner-Lamb-1977},
 \be\label{tcomp}
t_{\rm c}(r) = \frac{3 \pi\,r^2\,m_{\rm e}\,c^2}{2\,\sigma_{\rm T}\,L_{\rm X}},
 \ee
turns out to be less than the heating time due to magnetic field dissipation in case $L_{\rm X} > L_{\rm cr}$, where
 \be
 L_{\rm cr} \simeq 3 \times 10^{33}\ \mu_{30}^{1/4}\ m^{1/2}\ R_6^{-1/8}\ \left(\frac{\eta_{\rm m}}{0.001}\right) \left(\frac{R_{\rm sh}}{r_{\rm a}}\right)^{1/2}\ {\rm erg\,s^{-1}}.
 \ee
Here $m_{\rm e}$ is the electron mass, and $\sigma_{\rm T}$ is the Thompson cross-section. This implies that the process of magnetic accretion in 4U~2206+54 will take place according to scenario worked out by Bisnovatyi-Kogan and Ruzmaikin  \cite{Bisnovatyi-Kogan-Ruzmaikin-1974, Bisnovatyi-Kogan-Ruzmaikin-1976}, if $\eta_{\rm m} \la 0.1 (r_{\rm a}/R_{\rm sh})^{1/2}$. The neutron star magnetosphere in this case will be surrounded by the magnetic slab in which the radial motion will proceed as long as the magnetic field of the accretion flow dissipates. The density of plasma at the inner radius of the slab in the region of its interaction with the neutron star magnetosphere, $\rho_{\rm sl}$, can be evaluated from the balance of magnetic pressure and the thermal pressure in the slab as
 \be\label{rhosl}
  \rho_{\rm sl} = \frac{\mu^2\,m_{\rm p}}{2 \pi\,k_{\rm B}\,T_0\,r_{\rm m}^6}.
  \ee
Here $m_{\rm p}$ and $k_{\rm B}$  are the proton mass and Boltzmann constant, and $T_0$ is the gas temperature at the inner radius of the magnetic slab.

 \section{Spin evolution of the pulsar}

One of the signs of magnetic accretion in 4U~2206+54 is extremely high value of angular momentum dissipation in the accretion flow.
This conclusion is made taking into account that spin-down of neutron stars undergoing quasi-spherical accretion can occur provided the angular velocity of accreting material at the magnetospheric boundary, $\omega_{\rm em}(r_{\rm m}) = \xi \Omega_{\rm orb} \left(R_{\rm G}/r_{\rm m}\right)^2$, is less than the angular velocity of the star, $\omega_{\rm s}$ (see \cite{Bisnovatyi-Kogan-1991}). Solving inequality  $\omega_{\rm em}(r_{\rm m}) < \omega_{\rm s}$ for $\xi$, we find
  \begin{eqnarray}
\xi & < & 0.01~m^{-12/7}\ L_{35.6}^{-4/7}\ R_6^{20/7}\ \left(\frac{P_{\rm orb}}{19\,{\rm d}}\right) ~ \times\  \\
    \nonumber
 & & \times\  \left(\frac{P_{\rm s}}{5555\,{\rm s}}\right)^{-1} \left(\frac{v_{\rm rel}}{400\,{\rm km\,s^{-1}}}\right)^4 \left(\frac{B_*}{B_{\rm CRSF}}\right)^{8/7}.
  \end{eqnarray}
The derived value of $\xi$ is at least an order of magnitude less than the average value of this parameter evaluated in numerical simulations of the wind-fed accretion based on assumption  $\beta \gg 1$ (see \cite{Ruffert-1999} and references therein). So significant dissipation of the angular momentum in the accretion flow implies that either the neutron star is accreting material from the hot turbulent quasi-static envelope \cite{Davies-Pringle-1981, Ikhsanov-Beskrovnaya-2010}, or the infalling matter possesses high enough magnetic field and looses angular momentum due to magnetic viscosity \cite{Mestel-1959, Sparke-1982}.

The spin-up torque, applied to the neutron star from the accretion flow, $K_{\rm su} = \xi \Omega_{\rm orb} R_{\rm G}^2 \dot{M}$,
 \be
 K_{\rm su} \la 10^{31}\ \xi_{0.01} m^2 \dot{M}_{15} \left(\frac{P_{\rm orb}}{19\,{\rm d}}\right) \left(\frac{v_{\rm rel}}{400\,{\rm km\,s^{-1}}}\right)^{-4}\ {\rm dyne\,cm},
 \ee
under these conditions turns out to be significantly less than the absolute value of spin-down torque evaluated from observations,
 \be
K_{\rm sd} \ga 2 \pi I \dot{\nu}_{\rm sd} \sim 6 \times 10^{31} I_{45} \left(\frac{|\dot{\nu}_{\rm sd}|}{2 \times 10^{-14}\,{\rm Hz\,s^{-1}}}\right)\ {\rm dyne\,cm}.
 \ee
This allows to consider the equation of spin evolution of an accreting neutron star, $I \dot{\omega} = K_{\rm su} - K_{\rm sd}$, in a simplified form $I \dot{\omega} \simeq K_{\rm sd}$.

The spin-down torque applied to the neutron star from the accretion flow can be expressed as $K_{\rm sd}^{\rm (t)} = k_{\rm t} \dot{M}_{\rm c} \omega_{\rm s} r_{\rm m}^2$. Substituting  $\dot{M}_{\rm c} = \mu^2/\left(2 GM_{\rm ns} r_{\rm m}^7 \right)^{1/2}$ we get \cite{Ikhsanov-Beskrovnaya-2012}
 \be\label{ksdt}
K_{\rm sd}^{\rm (t)} = \frac{k_{\rm t}\,\mu^2}{\left(r_{\rm m}\,r_{\rm cor}\right)^{3/2}},
 \ee
where $r_{\rm cor} = \left(GM_{\rm ns}/\omega_{\rm s}^2\right)^{1/3}$ is the corotation radius of the neutron star. The value of spin-down torque essentially depends on the magnetospheric radius. In case of accretion from a turbulent envelope the magnetospheric radius is limited as $r_{\rm m} \ga r_{\rm a}$ \cite{Arons-Lea-1976, Davies-Pringle-1981}, and this estimate of the neutron star spin-down rate proves to be insufficient to explain the observed spin-down rate of the pulsar 4U~2206+54  \cite{Ikhsanov-Beskrovnaya-2010}. In case of accretion from the magnetic slab the value of magnetospheric radius depends on the mechanism by which the accretion flow penetrates into the stellar magnetic field. If the process of penetration is governed by interchange instabilities of the magnetospheric boundary (Rayleigh-Tailor and Kelvin-Helmholtz instabilities), then, as in the previous case, the magnetospheric radius will be close to its canonical value, $r_{\rm a}$. Otherwise it proves to be significantly less. If the interchange instabilities of the magnetospheric boundary are suppressed, the plasma penetration into the stellar field takes place due to diffusion and magnetic reconnection at the rate \cite{Elsner-Lamb-1984}
 \be\label{dmfin-1}
 \dot{M}_{\rm in}(r_{\rm m}) = 4 \pi r_{\rm m} \delta_{\rm m} \rho_0 V_{\rm ff}(r_{\rm m}) = 4 \pi r_{\rm m}^{3/2} D_{\rm eff}^{1/2} \rho_0 V_{\rm ff}^{1/2}(r_{\rm m}).
  \ee
Here $\delta_{\rm m} = \left(D_{\rm eff}\ \tau_{\rm d}\right)^{1/2}$ is the thickness of diffusion layer on the magnetospheric boundary (magnetopause), $D_{\rm eff}$ is the effective diffusion coefficient and   $\rho_0$ is the plasma density at the boundary.
The characteristic time of plasma diffusion into the magnetic field, $\tau_{\rm d}$, is determined by the time necessary for plasma having penetrated into the field to leave the magnetopause moving along the magnetospheric field lines in the gravitational field of the neutron star. Under the conditions of interest this time is close to dynamic (free-fall) time at the magnetospheric boundary, i.e. $\tau_{\rm d} \sim t_{\rm ff}(r_{\rm m})$ \cite{Elsner-Lamb-1984}.

The radius of the neutron star magnetosphere in this case can be evaluated from the condition of stationarity of the accretion process, adopting the rate of plasma penetration into the magnetic field of the neutron star equal to the rate of mass capture from the stellar wind, and, correspondingly, to the rate of mass accretion onto the stellar surface. Taking $D_{\rm eff} = D_{\rm B}$,
where
 \be\label{dbohm}
D_{\rm B} = \alpha_{\rm B} \frac{c k_{\rm B} T_0}{2 e B(r_{\rm m})}
 \ee
is the Bohm diffusion coefficient, and solving the equation $\dot{M}_{\rm in}(r_{\rm mca}) = L_{\rm X} R_{\rm ns}/GM_{\rm ns}$, we find
 \be\label{rmb}
r_{\rm mca} \simeq 2 \times 10^8\,{\rm cm}\ \times \ \alpha_{0.1}^{2/13}\ \mu_{30}^{6/13}\ T_6^{-2/13}\ m^{5/13}\ L_{35.6}^{-4/13}\ R_6^{-4/13}.
  \ee
Here $e$ is the electron charge, $T_6 = T_0/10^6$\,K is the plasma temperature and $B(r_{\rm m})$ is the magnetic field strength in the magnetopause. $\alpha_{0.1}=\alpha_{\rm B}/0.1$ is the efficiency parameter normalized according to observational results and numerical simulations of solar wind penetration into the Earth magnetosphere(see \cite{Gosling-etal-1991, Paschmann-2008} and references therein).

Combining expressions~(\ref{ksdt}) and (\ref{rmb}) and solving inequality $2 \pi I \dot{\nu}_{\rm sd} \la K_{\rm sd}^{\rm (t)}(r_{\rm mca})$, we find
  \begin{eqnarray}
|\dot{\nu}_{\rm sd}| & \ga & 2 \times 10^{-14}\,{\rm Hz\,s^{-1}} ~ k_{\rm t}\ \alpha_{0.1}^{-3/13}\ m^{-14/13}\ I_{45}^{-1}\ T_6^{3/13}\ L_{35.6}^{6/13} \\
     \nonumber
 & & \times \ R_6^{57/13} \left(\frac{P_{\rm s}}{5555\,{\rm s}}\right)^{-1}  \left(\frac{B_*}{B_{\rm CRSF}}\right)^{17/13}.
  \end{eqnarray}
Thus, the observed spin-down rate of the neutron star in the X-ray pulsar 4U~2206+54 can be explained in terms of magnetic accretion scenario assuming that the magnetic field strength on the neutron star surface is close to the value based on  the analysis of X-ray spectra of this pulsar, i.e. $\sim B_{\rm CRSF}$.

 \section{Conclusions}

Our study of rapid spin-down of the neutron star in 4U~2206+54 together with presented earlier picture of spin evolution in another long-period X-ray pulsar GX\,301--2 \cite{Ikhsanov-etal-2012, Ikhsanov-Beskrovnaya-2012} prove that application of the magnetic accretion scenario to the modeling of mass transfer in high-mass X-ray binaries is rather promising. One of the most important conditions for applicability of this scenario is magnetization of optical components in these systems which are usually giants and super-giants of early spectral type. Recent results of spectral and spectropolarimetric observations of massive stars (see, e.g. \cite{Martins-etal-2010, Hubrig-etal-2011} and references therein) show that  strong magnetization is not unique among these objects. Magnetic fields strength in the range from hundreds to thousands Gauss  (and in some cases in excess of 10\,kG) have been measured in more than 20 early-type stars. The value of parameter $\beta$ in the stellar wind from these objects is of the order of unity on the scale of orbital separation in close binary systems \cite{Walder-etal-2012}. Another criterion for realization of magnetic accretion in a binary system is a moderate velocity of stellar wind satisfying the condition $v_{\rm cr} < v_{\rm rel} < v_{\rm mca}$.

Thus, the systems with magnetic accretion occupy intermediate position between the systems with Keplerian accretion disks ($v_{\rm rel} \la v_{\rm cr}$) and the systems in which accretion onto a neutron star magnetosphere occurs in a form of free-falling  quasi-spherical flow  ($v_{\rm rel} \ga v_{\rm mca}$). Finally, it should be noted that magnetic accretion scenario turns out to be the most effective in interpretation of long-period neutron stars whose magnetospheric radius is significantly less than the corotation radius. The spin-down mechanism based on matter outflow from the magnetospheric boundary (see, e.g. \cite{Lovelace-etal-1995}) cannot be effective for these stars, while the spin-down torque caused by magnetic viscosity in the accretion flow remains significant.

\begin{theacknowledgments}
This work was supported by the Program of Presidium of the Russian Academy of Sciences N\,21, and NSH-1625.2012.2.
\end{theacknowledgments}


\begin{thebibliography}{99}

\bibitem[Giacconi et~al.(1972)]{Giacconi-etal-1972}
 Giacconi, R., Murray, S., Gursky, H., et~al., \emph{Astrophys. J.} \textbf {178}, 281 (1972).

\bibitem[Blay et~al.(2006)]{Blay-etal-2006}
 Blay, P., Negueruela, I., Rieg, P., et~al., \emph{Astron. Astrophys.}, \textbf{446}, 1095 (2006).

\bibitem[Corbet et~al.(2007)]{Corbet-etal-2007}
 Corbet, R.H.D., Markwardt, C.B., Tueller, J.,  \emph{Astrophys. J.} \textbf{655}, 458, (2007).

\bibitem[Ribo et~al.(2006)]{Ribo-etal-2006}
 Rib\'o, M., Negueruela, I., Blay, P., et~al., \emph{Astron. Astrophys.}, \textbf{449}, 687 (2006).

\bibitem[Blay \& Reglero(2011)]{Blay-Reglero-2011}
 Blay, P., Reglero, V., \emph{Soci\'et\'e Royale des Sciences de Li\'ege, Bulletin}, \textbf{80}, 634 (2011).

\bibitem[Rieg et~al.(2009)]{Rieg-etal-2009}
 Rieg, P., J.M.\,Torrej\'on, J.M., Negueruela, I., et~al., \emph{Astron. Astrophys.}, \textbf{494}, 1073 (2009).

\bibitem[Finger et~al.(2010)]{Finger-etal-2010}
 M.H.\,Finger, M.H., Ikhsanov, N.R., Wilson-Hodge, C.A., Patel, S.K., \emph{Astrophys. J.} \textbf{709}, 1249 (2010).

\bibitem[Wang(2010)]{Wang-2010}
 Wang, W., \emph{Astron. Astrophys.}, \textbf{520}, 22 (2010).

\bibitem[Masevich \& Tutukov(1988)]{Masevich-Tutukov-1988}
 Masevich, A.G., Tutukov, A.V., ``Stellar evolution: theory and observations'', Nauka Publ., Moscow, (1988).

\bibitem[Ikhsanov \& Beskrovnaya(2010)]{Ikhsanov-Beskrovnaya-2010}
 Ikhsanov, N.R., Beskrovnaya, N.G., \emph{Astrophysics}, \textbf{53}, 237 (2010).

\bibitem[Torrej\'on et~al.(2004)]{Torrejon-etal-2004}
 Torrej\'on, J.M., Kreykenbohm, I., Orr, A., et~al., \emph{Astron. Astrophys.}, \textbf{423}, 301, (2004).

\bibitem[Ikhsanov(2007)]{Ikhsanov-2007}
 Ikhsanov, N.R., \emph{Mon. Not. R. Astron. Soc.}, \textbf{375}, 698 (2007).

\bibitem[Colpi et~al.(2000)]{Colpi-etal-2000}
 Colpi, M., Geppert, U., Page, D..  \emph{Astrophys. J.} \textbf {529}, L29 (2000).

\bibitem[Bonanno et~al.(2006)]{Bonanno-etal-2006}
 Bonanno, A., Urpin, V., Belvedere, G., \emph{Astron. Astrophys.}, \textbf{451}, 1049, (2006).

\bibitem[Lipunov(1982)]{Lipunov-1982}
 Lipunov, V.M., \emph{Astrophys. and Space Sci.}, \textbf{85}, 451 (1982).

\bibitem[Doroshenko et~al.(2010)]{Doroshenko-etal-2010}
 Doroshenko, V., Santangelo, A., Suleimanov, V., et~al., \emph{Astron. Astrophys.}, \textbf{515}, 10 (2010).

\bibitem[Ikhsanov et~al.(2012)]{Ikhsanov-etal-2012}
 Ikhsanov, N.R., Pustil'nik, L.A., Beskrovnaya, N.G., \emph{AIP Conference Proceedings}, {\bf 1439}, 237 (2012).

\bibitem[Ikhsanov \& Beskrovnaya(2012)]{Ikhsanov-Beskrovnaya-2012}
 Ikhsanov, N.R., Beskrovnaya, N.G., \emph{Astron. Rep.}, \textbf{56}, 589 (2012).

\bibitem[Shvartsman(1971)]{Shvartsman-1971}
Shvartsman, V.F., \emph{Sov. Astron.}, \textbf{15}, 377 (1971).

\bibitem[Bisnovatyi-Kogan and Ruzmaikin(1974)]{Bisnovatyi-Kogan-Ruzmaikin-1974}
 Bisnovatyi-Kogan, G.S., Ruzmaikin, A.A., \emph{Astrophys. and Space Sci.}, \textbf{28}, 45 (1974).

\bibitem[Bisnovatyi-Kogan and Ruzmaikin(1976)]{Bisnovatyi-Kogan-Ruzmaikin-1976}
  Bisnovatyi-Kogan, G.S., Ruzmaikin, A.A., \emph{Astrophys. and Space Sci.}, \textbf{42}, 401 (1976).

\bibitem[Bondi \& Hoyle(1944)]{Bondi-Hoyle-1944}
 Bondi, H., Hoyle, F., \emph{Mon. Not. R. Astron. Soc.}, \textbf{104}, 273 (1944).

\bibitem[Bondi(1952)]{Bondi-1952}
 Bondi, H., \emph{Mon. Not. R. Astron. Soc.}, \textbf{112}, 195 (1952).

\bibitem[Arons and Lea(1976)]{Arons-Lea-1976}
 Arons, J., Lea, S.M., \emph{Astrophys. J.}, \textbf{207}, 914 (1976).

\bibitem[Elsner and Lamb(1977)]{Elsner-Lamb-1977}
 Elsner R.F., Lamb F.K., \emph{Astrophys. J.}, \textbf{215}, 897 (1977).

\bibitem[Illarionov \& Sunyaev(1975)]{Illarionov-Sunyaev-1975}
 Illarionov, A.F., Sunyaev, R.A. \emph{Astron. Astrophys.} \textbf{39}, 185 (1975).
 
\bibitem[Ruffert(1999)]{Ruffert-1999}
 Ruffert, M., \emph{Astron. Astrophys.}, \textbf{346}, 861 (1999). 

\bibitem[Zeldovich and Shakura(1969)]{Zeldovich-Shakura-1969}
 Zel'dovich, Ya.B., Shakura, N.I., \emph{Soviet Astronomy}, \textbf{13}, 175 (1969).

\bibitem[Bisnovatyi-Kogan and Fridman(1970)]{Bisnovatyi-Kogan-Fridman-1970}
 Bisnovatyi-Kogan, G.S., Fridman, A.M., \emph{Soviet Astronomy}, \textbf{13}, 566 (1970).

\bibitem[Kadomtsev(1987)]{Kadomtsev-1987}
 Kadomtsev, B.B. \emph{Reports on Progress in Physics}, \textbf{50}, 115 (1987).

\bibitem[Igumenshchev et~al.(2003)]{Igumenshchev-etal-2003}
 Igumenshchev, I.V., Narayan, R., Abramowicz, M.A., \emph{Astrophys. J.}, \textbf{592}, 1042 (2003).

\bibitem[Igumenshchev(2006)]{Igumenshchev-2006}
 Igumenshchev, I.V., \emph{Astrophys. J.}, \textbf{649}, 361 (2006).

\bibitem[Bisnovatyi-Kogan(1991)]{Bisnovatyi-Kogan-1991}
  Bisnovatyi-Kogan, G.S., \emph{Astron. Astrophys.}, \textbf{245}, 528 (1991).

\bibitem[Davies and Pringle(1981)]{Davies-Pringle-1981}
 Davies, R.E., Pringle, J.E., \emph{Mon. Not. R. Astron. Soc.}, \textbf{196}, 209 (1981).

\bibitem[Mestel(1959)]{Mestel-1959}
 Mestel, L., \emph{Mon. Not. R. Astron. Soc.}, \textbf{119}, 223 (1959).

\bibitem[Sparke(1982)]{Sparke-1982}
 Sparke, L.S., \emph{Astrophys. J.}, \textbf{260}, 104 (1982).
 
\bibitem[Elsner and Lamb(1984)]{Elsner-Lamb-1984}
 Elsner R.F., Lamb F.K., \emph{Astrophys. J.}, \textbf{278}, 326 (1984). 
 
\bibitem[Gosling et~al.(1991)]{Gosling-etal-1991}
  Gosling J.T., Thomsen M.F., Bame S.J., et~al., 1991, \emph{J. Geophys. Res.}, \textbf{96}, 14097  (1991).
  
\bibitem[Paschmann(2008)]{Paschmann-2008}
 Paschmann, G., \emph{Geophys. Res. Lett.}, \textbf{35}, L19109 (2008).   
 
\bibitem[Martins et~al.(2010)]{Martins-etal-2010}
  Martins, F., Donati, J.-F., Marcolino, W.L.F., et~al., \emph{Mon. Not. R. Astron. Soc.}, \textbf{407}, 1423 (2010). 

\bibitem[Hubrig et~al.(2011)]{Hubrig-etal-2011}
 Hubrig, S., Sch\"oller, M., Kravchenko, N.V., et~al., \emph{Astron. Astrophys.}, \textbf{582}, 151  (2011).

\bibitem[Walder et~al.(2012)]{Walder-etal-2012}
Walder, R., Folini, D., Meynet, G., \emph{Space Sci. Rev.}, \textbf{166}, 145 (2012).

\bibitem[Lovelace et~al.(1995)]{Lovelace-etal-1995}
 Lovelace, R.V.E., Romanova, M.M., Bisnovatyi-Kogan, G.S. \emph{Mon. Not. R. Astron. Soc.}, \textbf{275}, 244 (1995).

\end{thebibliography}
\end{document}